\RequirePackage{fix-cm}
\documentclass[smallextended]{svjour3}       
\smartqed  
\usepackage{graphicx}
\usepackage{amssymb}
\usepackage{amsmath}
\begin{document}

\title{Gravitational Collapse, Shear-free Anisotropic Radiating Star}\thanks{Grants or other notes
}


\author{B.C. Tewari \and
   Kali Charan
}


\institute{B.C. Tewari\and
           K. Charan \at
              Department of Mathematics, Kumaun University, SSJ Campus, Almora, Uttarakhand-263601,  India. \\
\vspace{.1 cm}
              \email drbctewari@yahoo.co.in\\
    \vspace{.1 cm}
               K. Charan\\
               email: kcyadav2008@gmail.com
}


\maketitle

\begin{abstract}
\noindent We here present a new model of a radiating star for a shear-free spherically symmetric anisotropic fluid undergoing radial heat flow collapsing under its own gravity. The interior metric fulfilled all the relevant physical and thermodynamic conditions and matched with Vaidya exterior metric over the boundary. Initially the interior solutions represent a static configuration of perfect fluid which then gradually starts evolving into radiating collapse. We have observed that the model is well behaved for a set of model parameters and there are a number of such sets for which the model is well behaved. The apparent luminosity as observed by the distant observer at rest at infinity and the effective surface temperature are zero in remote past at the instant when the collapse begins and at the stage when collapsing configuration reaches the horizon of the black hole.

\keywords{Exact solutions \and Anisotropic radiating star \and Gravitational collapse\and Black hole}
\end{abstract}

\section{Introduction}	
The study of gravitational collapse attracts researchers towards it by the fact that it represents theoretical and observable phenomena in the universe. It is an important open issue in relativistic astrophysics whether the final outcome of the gravitational collapse is formation of a black hole or a naked singularity(Joshi-Malafarina [1] and references therein).  Still there is no established theory available which can determine the final fate of this collapse. There is no iron-clad evidence that black hole candidates are indeed black holes. There is no logic that prevents existence of naked singularities and as per Cosmic Censorship Conjecture Penrose [2] himself considers this an open question. Understanding the characteristics and features of final fate of a collapsing system is not just important from the theoretical perspective; it has tremendous observational consequences as well. Virbhadra with his collaborators showed that Black holes and naked singularities could be observationally differentiated through their gravitational lensing features (Virbhadra-Ellis [3,4]; Virbhadra-Keeton [5]; Virbhadra [6] and also references therein).

 In view of classical gravity, to understand the nature of collapse and physical behaviour of a collapsing system properly, it is necessary to construct a realistic model of the collapsing system. It is very difficult task on account of the highly non linear nature of the governing field equations. Many efforts have been made for reducing the complexity and various methods for simplifications are frequently adopted and the pioneering work of Oppenheimer-Snyder [7] was the initial step in this direction when collapse of a highly idealised spherically symmetric dust cloud was studied. Since then several attempts have been made to construct realistic models of gravitationally collapsing systems to understand the properties and nature of collapsing objects. It gained to tremendous momentum when Vaidya [8] presented a solution describing the exterior gravitational field of a stellar body with outgoing radiation and the modified equations in this case proposed by Misner [9] and Lindquist et al.[10] for an adiabatic distribution of matter.

 It is an established fact that gravitational collapse is a highly dissipating energy process (Herrera-Santos [11]; Herrera et al. [12], Mitra [13],[14]). However, the dissipation of energy from collapsing fluid distribution is described in two limiting cases. The first case describes the free streaming approximation and a number of solutions of radiating fluid ball in this case discussed by Tewari [15-18]. Further Pant- Tewari [19] presented a quasar model with all degree of suitability. While second one is diffusion approximation and in this case the dissipation is modeled by heat flow type vector and in this the model proposed by Glass [20] has been extensively studied by Santos [21] for the junction conditions of collapsing spherically symmetric shear-free non-adiabatic fluid with radial heat flow. On a similar ground a number of studies have been reported by de Oliveira et al. [22]; de Oliveira-Santos [23]; Bonnor et al. [24]; Banerjee et al. [25]; Govinder-Govender [26]; Maharaj et al. [27]; Herrera et al.[28-30]; Naidu-Govender [31]; Sarwe-Tikekar [32]; Ivanov [33]; Pinheiro-Chan [34]; Tewari [35, 36]; Tewari-Charan [37, 38] and also references therein for describing a collapsing fluid radiating energy. A remarkable work for collapsing anisotropic radiating star is due to Herrera -Santos [39] explored the properties of anisotropic self-gravitating spheres using the perturbation method. Herrera [40] investigated that the local pressure anisotropy is one of the responsible factors for inhomogeneities in energy density. On a same ground a  noticeable work is also due to Tikekar and Sharma [41]; Nguyen-Pedraza [42]; Nguyen-Lingam [43]; Sharif-Abbas [44]; Abbas [45]; Tewari-Charan [46], Ivanov ([47], [48]) and also references therein have been reported with the impact of inhomogeneity, anisotropy and various dissipative processes on the evolution.

The main objective of this work is to shed some light on the final fate of gravitational collapse by taking Tewari and Charan [46] solution as seed solution. Final fate of our model is the formation of a black hole. By taking a set of model parameters we have developed a Supernovae model. The interior matter fluid is spherically symmetric, shear-free, anisotropic radiating away its energy in the form of radial heat and contracting in size during the process of collapse. The interior metric matched with the Vaidya exterior metric [8] over the boundary. The paper is organised as: In Section 2 the interior space-time and related equations for collapsing system are given. In Section 3 the exterior space-time and junction conditions  are presented for matching the two space-times. Section 4 describes a detailed study of the model in which we have obtained expressions of various physical parameters and temperature profile of the collapsing body. Finally in Section 5 some concluding remarks have been made.
\section{The interior space-time and related equations for collapsing system}
The interior metric  of a shear-free spherically symmetric fluid  is given by
\begin{equation}
ds_-^2=-A^2(r,t)dt^2+B^2(r,t)\{dr^2+r^2(d\theta^2+\sin^2\theta d\phi^2)\}
\end{equation}
The energy-momentum tensor for the matter distribution with anisotropy in pressure is
\begin{equation}
T_{\mu \nu}=(\epsilon +p_t)w_\mu w_\nu+p_t g_{\mu\nu}+(p_r-p_t)x_\mu x_\nu+q_\mu w_\nu+q_\nu w_\mu
\end{equation}
where $\epsilon$ is the energy density of the fluid, $p_r$ the radial pressure,  $p_t$  the tangential pressure, $w_\mu$ is the four- velocity, $q_\mu$ the radial heat flow vector and  $x_\mu$ is a unit space like four vector along the radial direction.

 Assuming comoving coordinates, we have
$w^\mu=\frac{1}{A}\delta_0^\mu$.
The heat flow vector $q^\mu$ is orthogonal to the velocity vector so that $q^\mu w_\mu=0$ and takes the form
$q^\mu=q \delta_1^\mu$.The line element (1) corresponds to shear- free spherically symmetric fluid (Glass [49]), as the shear tensor vanishes identically.

In order to solve the Non-trivial Einstein’s field equations which are generated by (1) and (2), we choose a particular form of the metric coefficients given in (1) into functions of $r$ and $t$ coordinates as $A(r,t)=A_0(r)g(t)$ and $B(r,t)=B_0(r)f(t)$.  The coupling constant in geometrized units is taken as $\kappa=8\pi  (i.e.  G = c = 1)$ and in view of (1) and (2) with the help of above metric coefficients we get the following expressions for field equations
 \begin{eqnarray}
&\kappa\epsilon=\frac{\epsilon_0}{f^2}+\frac{3\dot{f}^2}{A_0^2g^2f^2}\\
&\kappa p_r=\frac{(p_r)_0}{f^2}+\frac{1}{A_0^2g^2}\Big(-\frac{2\ddot{f}}{f}-\frac{\dot{f}^2}{f^2}+\frac{2\dot{f}\dot{g}}{fg}\Big)\\
&\kappa p_t=\frac{(p_t)_0}{f^2}+\frac{1}{A_0^2g^2}\Big(-\frac{2\ddot{f}}{f}-\frac{\dot{f}^2}{f^2}+\frac{2\dot{f}\dot{g}}{fg}\Big)\\
&\kappa q=-\frac{2A_0'\dot{f}}{A_0^2B_0^2gf^3}
\end{eqnarray}
where
\begin{eqnarray}
&\epsilon_0=-\frac{1}{B_0^2}\Big(\frac{2B_0''}{B_0}-\frac{B_0'^2}{B_0^2}+\frac{4B_0'}{rB_0}\Big)\\
&(p_r)_0=\frac{1}{B_0^2}\Big(\frac{B_0'^2}{B_0^2}+\frac{2B_0'}{rB_0}+\frac{2A_0'B_0'}{A_0B_0}+\frac{2A_0'}{rA_0}\Big)\\
&(p_t)_0=\frac{1}{B_0^2}\Big(\frac{B_0''}{B_0}-\frac{B_0'^2}{B_0^2}+\frac{B_0'}{rB_0}+\frac{A_0''}{A_0}+\frac{A_0'}{rA_0}\Big)
\end{eqnarray}
Here the quantities with the suffix 0 corresponds to the static star model with metric components $A_0(r),~B_0(r)$ and the primes and dots stand for differentiation with respect to $r$ and $t$ respectively.

 To find a new parametric class of exact solutions of pressure anisotropy equation which is created by the equations (4) and (5),  Tewari-Charan [46] assumed that the anisotropy evolves
$\kappa( p_t-p_r)=\Delta(r,t)=\frac{\delta(r)}{B_0^2 (r) f^2 (t)}$ and  got the following  time independent differential equation
\begin{eqnarray}
\frac{A_0''}{A_0}+\frac{B_0''}{B_0}-\frac{A_0'}{rA_0}-\frac{B_0'}{rB_0}-\frac{2B_0'^2}{B_0^2}-\frac{2A_0'B_0'}{A_0B_0}-\delta(r)=0
\end{eqnarray}
Making an adhoc relationship between the variables in (10), Tewari- Charan [46] obtained the following solution
\begin{eqnarray}
& A_0=C_4(1+C_3r^2)(1+C_1r^2)^{\frac{n}{l+1}}\\
& B_0=C_2(1+C_1r^2)^{\frac{1}{l+1}}\\
& \delta(r)=\frac{(2n-2)}{(l+1)}\frac{4C_3C_1r^2}{(1+C_3r^2)(1+C_1r^2)}
\end{eqnarray}
where $n,l,C_1,C_2,C_3$  and $C_4$ are constants and
\begin{eqnarray}
n=\frac{1}{2}\Big\{(l+3)\pm(l^2+10l+17)^{\frac{1}{2}}\Big\}
\end{eqnarray}
where $n$ is real if $l\geq -5+2\sqrt{2}$   or  $l\leq-5-2\sqrt{2}$.

\section{The exterior space-time and  junction conditions }
 The exterior space-time of a collapsing radiating star is described by Vaidya exterior metric [8]

\begin{equation}
ds_+^2=-\Big(1-\frac{2M(v)}{R}\Big)dv^2-2dRdv+R^2(d\theta^2+\sin^2\theta d\phi^2)
\end{equation}

\noindent where $v$ is the retarded time and  $M(v)$ is the exterior Vaidya mass.

\vspace{.1cm}
\noindent The junction conditions for radiating star matching two line elements (1) and (15) at the boundary continuously across a spherically symmetric time-like hyper surface $\Sigma$ are very well known and given by Santos [21]
\begin{eqnarray}
&(rB)_\Sigma=R_\Sigma(v)= \mathcal{R}(\tau)\\
&(p_r)_\Sigma= (qB)_\Sigma\\
&m_\Sigma(r,t)= M(v)=\Bigg\{\frac{r^3B\dot{B}^2}{2A^2}-r^2B'-\frac{r^3B'^2}{2B}\Bigg\}_\Sigma
\end{eqnarray}
where $m_\Sigma$ is the mass function calculated in the interior at $r=r_\Sigma$ (Cahill et al. [50]; Misner-Sharp [51]).

\vspace{.1cm}

\noindent  The surface luminosity and the boundary redshift $z_\Sigma$ observed on $\Sigma$ are obtained by de Oliveira et al. [22]
\begin{eqnarray}
& L_\Sigma=\frac{\kappa}{2}\{r^2B^3q\}_\Sigma\\
& z_\Sigma=\Big[1+\frac{rB'}{B}+\frac{r\dot{B}}{A}\Big]_\Sigma^{-1}-1
\end{eqnarray}
The total luminosity for an observer at rest at infinity is
\begin{eqnarray}
L_\infty=-\frac{dM}{dv}=\frac{L_\Sigma}{(1+z_\Sigma)^2}
\end{eqnarray}
	
 In the absence of non-adiabatic dissipative forces the equation (17), $(p_r)_\Sigma =(qB)_\Sigma$, reduces to the condition $[(p_r )_0]_\Sigma= 0$ and yields at $r=r_\Sigma=R_\Sigma$

\begin{eqnarray}
\frac{2\ddot{f}}{f}+\frac{\dot{f}^2}{f^2}-\frac{2\dot{g}\dot{f}}{gf}=\frac{2\alpha g\dot{f}}{f^2}
\end{eqnarray}
where
\begin{eqnarray}
\alpha=\Big(\frac{A_0'}{B_0}\Big)_\Sigma
\end{eqnarray}
Motivated by de Oliveira et al. [22] and Bonnor et al. [24], Tewari [36] assume $g(t)=f(t)$ and got the solution of (22). The solution so obtained is identical to the solution presented by de Oliveira et al. [22]; Bonnor et al. [24] with $g(t)=1$. Keeping in mind the collapsing configuration range finally Tewari [36] obtained
\begin{eqnarray}
&\dot{f}=-2\alpha \sqrt{f}(1-\sqrt{f})\\
&t=\frac{1}{\alpha}ln(1-\sqrt{f})
\end{eqnarray}
We observed that the function $f(t)$ decreases monotonically from the value $f(t)=1$ at $t=-\infty$ to $f(t)=0$ at $t=0$. It interpret that the collapse of the configuration begins in the remote past and gradually starts evolve into radiating collapse.

\section{Detailed study of general model of collapsing radiating star}
Using (3)-(9), (11),(12) and (24)and (25), the general expressions for the energy density, both the pressures and the heat flux for the collapsing body are
\begin{eqnarray}
&\kappa\epsilon=\frac{\epsilon_0}{f^2}+\frac{12\alpha(1-\sqrt{f})^2}{f^3\Big[C_4(1+C_3r^2)(1+C_1r^2)^{\frac{n}{l+1}} \Big]^2}\\
&\kappa p_r=\frac{(p_r)_0}{f^2}+\frac{4\alpha^2(1-\sqrt{f})}{f^{\frac{5}{2}}\Big[C_4(1+C_3r^2)(1+C_1r^2)^{\frac{n}{l+1}}\Big]^2}\\
&\kappa p_t=\frac{(p_t)_0}{f^2}+\frac{4\alpha^2(1-\sqrt{f})}{f^{\frac{5}{2}}\Big[C_4(1+C_3r^2)(1+C_1r^2)^{\frac{n}{l+1}}\Big]^2}\\
&\kappa q =\frac{4r}{(l+1)C_4C_2^2(1+C_3r^2)^2(1+C_1r^2)^{\frac{n+2}{l+1}+1}}[(l+1)C_3(1+C_1r^2) \nonumber \\
&~~~~~~~~~~~~~~~~~~~~~~~~~~~~~~~~~~~~~~~~~~~~~ +nC_1(1+C_3r^2)]\frac{2\alpha(1-\sqrt{f})}{f^{\frac{7}{2}}}
\end{eqnarray}

where $\epsilon_0,(p_r)_0,(p_t)_0$  are energy density, radial and tangential pressures respectively in static position of the star and they are given by
\begin{eqnarray}
&\epsilon_0=\frac{4C_1}{(l+1)^2C_2^2(1+C_1r^2)^{\frac{2}{l+1}+2}}[-3(l+1)-(l+2)C_1r^2]\\
&(p_r)_0=\frac{4C_1}{(l+1)^2C_2^2(1+C_1r^2)^{\frac{2}{l+1}+2}}[(l+1)(n+1)+n^2C_1r^2 \nonumber \\
&~~~~~~~~~~~~~~~~~~~~~~~~~~~~~~~~~~~~~~~~~~+\frac{(l+1)C_3(1+C_1r^2)\{(l+1)+(l+3)C_1r^2\}}{C_1(1+C_3r^2)}]\\
&(p_t)_0=\frac{4C_1}{(l+1)^2C_2^2(1+C_1r^2)^{\frac{2}{l+1}+2}}[(l+1)(n+1)+n^2C_1r^2 \nonumber \\
&~~~~~~~~~~~~~~~~~~~~~~~~~~~~~~~~~~~ ~~ +\frac{(l+1)C_3(1+C_1r^2)\{(l+1)+(l+2n+1)C_1r^2\}}{C_1(1+C_3r^2)}]
\end{eqnarray}

By utilizing (11), (12) and (23), we have
\begin{eqnarray}
&\alpha =\frac{2C_4r_\Sigma(1+C_1r_\Sigma^2)^{\frac{n-l-2}{l+1}}}{C_2(l+1)}[nC_1(1+C_3r_\Sigma^2)+(l+1)C_3(1+C_1r_\Sigma^2)]
\end{eqnarray}
\subsection{Detailed study of a specific model}
For different values of $n$, one can obtained a number of solutions using equations (11) and (12). For $n=0$, we get the homogeneous density and anisotropic pressure, for $n=1$, we get the homogeneous density and isotropic pressure and for $n=-1$, we get the isotropic pressure. In this way a number of different radiating star models can be develop from the above obtained solution. In order to maintain the inhomogeniety and anisotropy a Horizon-free case for $n=-1-\sqrt{2}$ has been studied by Tewari-Charan [48]. In the present study we assume $n=-2$, to maintain the inhomogeniety and anisotropy for a collapsing radiating star, so utilizing (11)-(14), (30)-(32) we get
\begin{eqnarray}
&A_0=C_4(1+C_3r^2)(1+C_1r^2)^{\frac{2}{7}}\\
& B_0=C_2(1+C_1r^2)^{\frac{-1}{7}}\\
& \delta(r)=\frac{24}{7}\frac{C_3C_1r^2}{(1+C_3r^2)(1+C_1r^2)}\\
&\epsilon_0=\frac{4C_1}{49C_2^2(1+C_1r^2)^{\frac{12}{7}}}[21+6C_1r^2]\\
&(p_r)_0=\frac{4C_1}{49C_2^2(1+C_1r^2)^{\frac{12}{7}}}[(7+4C_1r^2)+\frac{7C_3(1+C_1r^2)(7+5C_1r^2)}{C_1(1+C_3r^2)}]\\
&(p_t)_0=\frac{4C_1}{49C_2^2(1+C_1r^2)^{\frac{12}{7}}}[(7+4C_1r^2)+\frac{7C_3(1+C_1r^2)(7+11C_1r^2)}{C_1(1+C_3r^2)}]
\end{eqnarray}
The junction condition $\{(p_r)_0\}_\Sigma=0$ gives
\begin{eqnarray}
C_3=\frac{-C_1(7+4C_1r_\Sigma^2)}{7(1+C_1r_\Sigma^2)(7+5C_1r_\Sigma^2)+C_1r_\Sigma^2(7+4C_1r_\Sigma^2)}
\end{eqnarray}
Here from (38) and (39), we are seeing that at the centre radial and tangential pressures are equal and anisotropy vanishes there.

 A physically reasonable solution should satisfy certain conditions and which are:
 
 The central values of both the pressures, density and gravitational potential component should be non-zero positive definite. This condition gets satisfied if $C_1>0,C_2>0,C_4>0$ and $C_3>(- C_1)/7$ , subjecting to the condition $\frac{(p_r)_0}{\epsilon_0} <1$ which yields $C_3<\frac{2C_1}{7}$.
  
Secondly the solution should have monotonically decreasing expressions for both the pressures and density with the increase of radial coordinate $r$. In view of this we have observed that these conditions hold good if $C_1>0$ and  $C_3<0$ satisfying all other requirement of the model.
Further, it is mentioned here that the boundary of the collapsing radiating star is established only when $C_1>0,C_2>0,C_4>0$ and $\frac{- C_1}{7}<C_3<0$ .

Now utilizing (26)-(29), (34) and (35) the expressions for $\epsilon, p_r,p_t,q$  reduce to the following
\begin{eqnarray}
&\kappa\epsilon=\frac{\epsilon_0}{f^2}+\frac{12\alpha^2(1-\sqrt{f})^2}{f^3\Big[C_4(1+C_3r^2)(1+C_1r^2)^{\frac{2}{7}} \Big]^2}\\
&\kappa p_r=\frac{(p_r)_0}{f^2}+\frac{4\alpha^2(1-\sqrt{f})}{f^{\frac{5}{2}}\Big[C_4(1+C_3r^2)(1+C_1r^2)^{\frac{2}{7}}\Big]^2}\\
&\kappa p_t=\frac{(p_t)_0}{f^2}+\frac{4\alpha^2(1-\sqrt{f})}{f^{\frac{5}{2}}\Big[C_4(1+C_3r^2)(1+C_1r^2)^{\frac{2}{7}}\Big]^2}\\
& \kappa q=\frac{4r}{7C_4^2C_2^2(1+C_3r^2)^2(1+C_1r^2)}[2C_1+C_3(7+9C_1r^2)]\frac{2\alpha(1-\sqrt{f})}{f^{\frac{7}{2}}}\
\end{eqnarray}
The fluid collapse rate $\Theta =w_{;\mu}^\mu $  with the help of (1), (24), (34) and (35) is 
\begin{equation}
\Theta=\frac{-6\alpha(1-\sqrt{f})}{f^{\frac{3}{2}}\Big[C_4(1+C_3r^2)(1+C_1r^2)^\frac{2}{7}\Big]}
\end{equation}
where by using (33)
\begin{eqnarray}
\alpha =\frac{2C_4r_\Sigma}{7C_2(1+C_1r_\Sigma^2)^{\frac{4}{7}}}[2C_1+C_3(7+9C_1r_\Sigma^2)]
\end{eqnarray}
To develop a Supernovae model and pictorial representation of various physical parameters keeping in mind the reasonable conditions we have chosen the model parameters as $C_1 = 1$, $C_2 = 0.1$, $M_0 = 5M_\Theta$, $C_4 = 0.5$
\begin{figure}[ht]
\begin{center}
\includegraphics[scale=0.6]{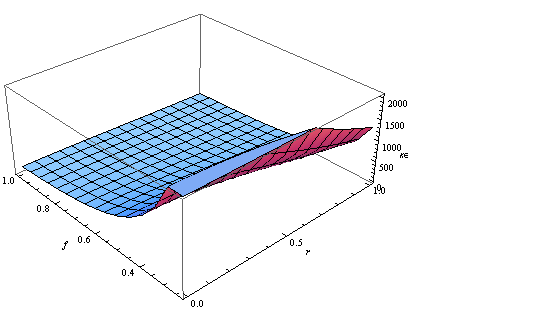}
\caption{Behavior of density vs radial and temporal co-ordinates}
\end{center}
\end{figure}
\begin{figure}
\begin{center}
\includegraphics[scale=0.6]{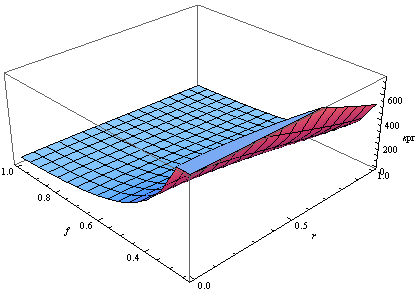}
\caption{Behavior of radial pressure vs radial and temporal co-ordinates}
\includegraphics[scale=0.6]{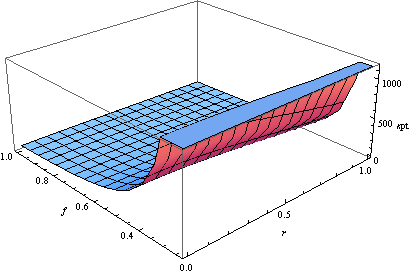}
\caption{Behavior of tangential pressure vs radial and temporal co-ordinates}
\includegraphics[scale=0.6]{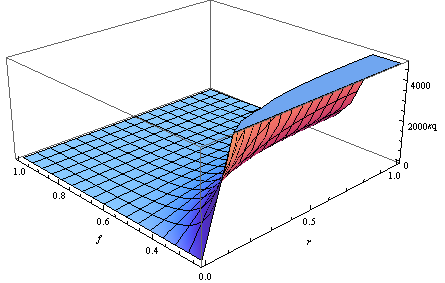}
\caption{Behavior of heat flux vs radial and temporal co-ordinates}
\end{center}
\end{figure}
\begin{figure}
\begin{center}
\includegraphics[scale=0.6]{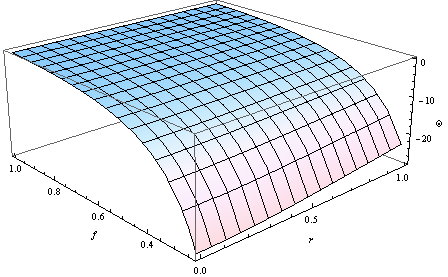}
\caption{Behavior of collapse rate vs radial and temporal co-ordinates}
\end{center}
\end{figure}
\newpage
\noindent
We can see [fig.1-4] the physical parameters  are finite, positive, monotonically decreasing at any instant with respect to radial coordinate for $0 \leq r \leq r_\Sigma $. The plots of various physical parameters  show that the nature of these parameters is well behaved with respect to radial and temporal coordinates for a set of constraints and there are a number of such constraints for which the model is well behaved. Fig. 5 shows that initially collapse rate is zero and it increases with respect to radial and temporal coordinates till the formation of black hole. 
\subsection{Mass energy, physical radius and time of black hole formation}
The total mass energy entrapped inside the surface $\Sigma$ is given by (18), which becomes by using (24), (34), (35) and (46)
\begin{eqnarray}
M(v)=\Bigg[\frac{8C_1^2C_2r_\Sigma^5[2C_1+C_3(7+9C_1r_\Sigma^2)]^2(1-\sqrt{f})^2}{49(1+C_1r_\Sigma^2)^{\frac{15}{7}}(1+C_3r_\Sigma^2)^2}+m_0f\Bigg]_\Sigma
\end{eqnarray}
where
\begin{eqnarray}
m_0=\frac{2C_1C_2r_\Sigma^3(7+6C_1r_\Sigma^2)}{49(1+C_1r_\Sigma^2)^{\frac{15}{7}}}
\end{eqnarray}
Using (16) and (35), we get the physical radius of the collapsing radiating star as
\begin{eqnarray}
 R_\Sigma(v)=r_\Sigma C_2(1+C_1r_\Sigma^2)^{-1/7}f
\end{eqnarray}
\begin{figure}[ht]
\begin{center}
\includegraphics[scale=0.6]{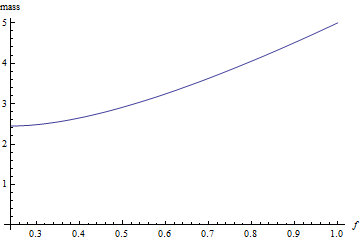}
\caption{Behavior of mass vs temporal co-ordinate}
\includegraphics[scale=0.6]{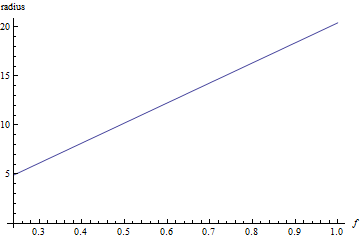}
\caption{Behavior of radius vs temporal co-ordinate}
\end{center}
\end{figure}

Utilizing (19)-(21), (24), (34), (35) and (46) the surface luminosity, the boundary redshift on $\Sigma$  and the luminosity for distant observer at rest at infinity are
\begin{eqnarray}
& L_\Sigma=\frac{8}{49}\Bigg[ \frac{C_1r_\Sigma^2[2C_1+C_3(7+9C_1r_\Sigma^2)]}{(1+C_1r_\Sigma^2)(1+C_3r_\Sigma^2)}\Bigg]^2\frac{(1-\sqrt{f})}{\sqrt{f}}\\
& z_\Sigma=\Bigg[ \frac{7+5C_1r_\Sigma^2}{7(1+C_1r_\Sigma^2)}-\frac{4r_\Sigma^2[2C_1+C_3(7+9C_1r_\Sigma^2)]}{(1+C_3r_\Sigma^2)(1+C_1r_\Sigma^2)}\frac{(1-\sqrt{f})}{\sqrt{f}}\Bigg]_\Sigma^{-1}-1
\end{eqnarray}
and
\begin{eqnarray}
L_\infty=\frac{8}{49}\Bigg[ \frac{C_1r_\Sigma^2[2C_1+C_3(7+9C_1r_\Sigma^2)]}{(1+C_1r_\Sigma^2)(1+C_3r_\Sigma^2)}\Bigg]^2\frac{(1-\sqrt{f})}{\sqrt{f}}\frac{1}{(1+z_\Sigma)^2}
\end{eqnarray}
The expression (52) shows that $L_\infty$ vanishes in the beginning when $f(t)\to 1$ and at the stage when $[z_\Sigma \to \infty]$.
The black hole formation time when the collapse reaches the horizon of the black hole occurs when the surface redshift goes to infinity, for this the term in the parentheses of (51) goes to zero and we obtain

\begin{eqnarray}
\sqrt{f}_{BH}=\frac{4C_1r_\Sigma^2[2C_1+C_3(7+9C_1r_\Sigma^2)]}{4C_1r_\Sigma^2[2C_1+C_3(7+9C_1r_\Sigma^2)]+(7+5C_1r_\Sigma^2)(1+C_3r_\Sigma^2)}
\end{eqnarray}
and
\begin{eqnarray}
t_{BH}=\frac{1}{\alpha}ln \Bigg[ \frac{(7+5C_1r_\Sigma^2)(1+C_3r_\Sigma^2)}{4C_1r_\Sigma^2[2C_1+C_3(7+9C_1r_\Sigma^2)]+(7+5C_1r_\Sigma^2)(1+C_3r_\Sigma^2)} \Bigg]
\end{eqnarray}
Figs. (6 and 7) show the behavior of mass energy and physical radius of collapsing radiating star. Initially when $f(t)\to1$ both the mass and physical radius of collapsing body are 5 ${M_\Theta}$  and 20.4 km  and at the time of black hole formation they remain 2.4 ${M_\Theta}$ and 4.86 km respectively. In view of (53) the time of black hole formation is observed as 0.2399 S.

\subsection{Temperature profile for collapsing radiating star}
To obtain the temperature inside and on the surface, we utilize temperature gradient law governing the heat transport within the collapsing matter (Israel et al. [52]; Maartens [53]; and Martinez [54]) given as
\begin{eqnarray}
\tau(g^{\mu \nu}+w^\mu w^\nu)w^\alpha q_{\nu ; \alpha}+q^\mu = -\mathbb{K}(g^{\mu \nu}+w^\mu w^\nu)[T_{,\nu}+T\dot{w_\nu}]
\end{eqnarray}
where $\mathbb{K}$ is the thermal conductivity and $\tau$ is the relaxation time. 

To get a simple estimate of the temperature evolution, we set relaxation time as zero in (55) and get
\begin{eqnarray}
q=-\mathbb{K}\frac{1}{B_0^2f^2}\Big(T'+T\frac{A_0'}{A_0}\Big)=-\frac{2A_0'\dot{f}}{A_0^2B_0^2f^4}
\end{eqnarray}
If we assume thermal conductivity  $\mathbb{K}=\gamma T^\Omega \geq 0$, where $\gamma$ and $\Omega$ are positive constants, then on integration (56) we get
\begin{eqnarray}
T^{\Omega+1}=\frac{T_0(t)}{A_0^{\Omega+1}}-\frac{4(\Omega+1)}{\gamma \Omega}\frac{\alpha}{A_0}\frac{1-\sqrt{f}}{f^{3/2}}
\end{eqnarray}
where $T_0 (t)$ is an arbitrary function of $t$.

 The effective surface temperature observed by external observer can be calculated from the following expression given by Schwarzschild [55] as
\begin{equation}
\begin{split}
T_\Sigma^4  &= \Big\{\frac{1}{\pi\delta(rB_0f)^2}\Big\}_\Sigma L_\infty \\
& =\frac{8}{49}\frac{C_1^2r_\Sigma^2}{\pi\delta C_2^2(1+C_1r_\Sigma^2)^{\frac{12}{7}}(1+C_3r_\Sigma^2)^2}\frac{(1-\sqrt{f})}{f^{5/2}}\frac{1}{(1+z_\Sigma)^2}
\end{split}
\end{equation}
 where for photons the constant $\delta$ is given by $\delta=(\pi^2 \mathrm{k}^4)/(15\hbar^3 )$, here $\mathrm{k}$ and $\hbar$ denoting respectively Boltzmann and Plank constants.
 
Choosing $\Omega=3$ which represents radiation interaction with matter fluid through the diffusive approximation (Misner and Sharp [56]). The arbitrary function $T_0 (t)$ is determined by using (57) and (58) as
\begin{equation}
\begin{split}
T_0(t)&= \Bigg\{\frac{16\alpha}{3\mathrm{k}\gamma}\frac{\Big[C_4(1+C_3r^2)(1+C_1r^2)^{2/7}\Big]^3(1-\sqrt{f})}{f^{3/2}}\Bigg\}_\Sigma  \\
& +\Bigg\{\frac{2\alpha^2\Big[C_4(1+C_3r^2)(1+C_1r^2)^{2/7}\Big]^2}{\pi\delta r^2}\frac{(1-\sqrt{f})}{f^{5/2}}\Bigg\}_\Sigma\frac{1}{(1+z_\Sigma)^2}
\end{split}
\end{equation}

Temperature distribution throughout the interior of the collapsing body is
\begin{eqnarray}
&T^4=\Bigg[\frac{T_0(t)}{\Big[C_4(1+C_3r^2)(1+C_1r^2)^{2/7}\Big]^4} \nonumber \\
&~~~~~~~~~~~~~~~~~~~~~~~~~~~~~~-\frac{16\alpha}{3\gamma\Big[C_4(1+C_3r^2)(1+C_1r^2)^{2/7}\Big]}\frac{(1-\sqrt{f})}{f^{3/2}}\Bigg]
\end{eqnarray}\\
\begin{figure}[ht]
\begin{center}
\includegraphics[scale=00.6]{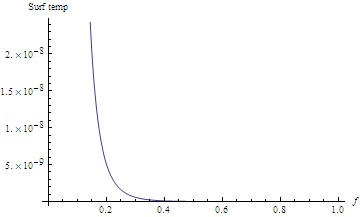}
\caption{Behavior of surface temperature vs temporal co-ordinate}
\end{center}
\end{figure}\\
It follows that the surface temperature of the collapsing star tends to zero at the beginning of the collapse $[f\to 1]$ and at the stage of formation of black hole $[z_\Sigma \to \infty]$, it is also evident from the fig. (8).

\section{Discussion and Conclusions}
We have given a new class of exact solutions of spherically symmetric shear-free anisotropic fluid distributions radiating away its energy in the form of radial heat flow. We have obtained a class of exact solutions by assigning different values to the parameter $n$. A simple radiating star model for $n=-2$  studied in detail. The model is physically and thermodynamically sound as it corresponds to well-behaved nature for the fluid density, both the radial and tangential pressures and the radiation heat flux throughout the fluid sphere. Initially the interior solutions represent a static configuration of dissipative fluid which then gradually starts evolving into radiating collapse. The apparent luminosity as observed by the distant observer at rest at infinity is zero in remote past at the instance when the collapse begins and at the stage of black hole formation.

We observed that the function $f(t)$ decreases monotonically from the value $f(t)=1$ at $t=-\infty$ to $f(t)=0$ at $t=0$. The plots of various physical parameters [figs.(1-4)] show that the nature of these parameters is well behaved with respect to radial and temporal coordinates for a set of constraints and there are a number of such sets for which the model is well behaved. Fig.(5) shows that initially collapse rate is zero and it increases with respect to radial and temporal coordinates till the formation of black hole. Figs.(6 and 7) show the behavior of mass energy and physical radius of collapsing radiating star. Initially when $f(t)\to 1$ both the mass and physical radius of a core of  pre Supernovae were 5 ${M_\Theta}$ and 20.4 km  and they remain 2.4 ${M_\Theta}$ and 4.86 km respectively at the time of black hole formation. The time of black hole formation is observed as $0.2399$ S. The surface temperature of the collapsing radiating star tends to zero at the beginning of the collapse $[f\to 1]$ and the stage of formation of black hole $[z_\Sigma \to \infty]$.

\vspace{.3cm}
\textbf{References}
\begin{enumerate}
\itemsep 1pt
\parskip 0pt
\item[1.]Joshi, P. S., Malafarina, D.: Int. J. Mod. Phys. D \textbf{20},  2641 (2011)
\item[2.]Penrose, R.: Riv. Nuovo Cimento. \textbf{1}, 252 (1969)
\item[3.] Virbhadra, K. S., Ellis, G. F. R.: Phys. Rev. D \textbf{62}, 084003 (2000)
\item[4.]Virbhadra, K. S., Ellis, G. F. R.: Phys. Rev. D \textbf{65},  103004 (2002)
\item[5.]Virbhadra, K. S., Keeton, C. R.: Phys. Rev. D \textbf{77}, 124014 (2008)
\item[6.]Virbhadra, K. S.: Phys. Rev. D \textbf{79}, 083004 (2009)
\item[7.] Oppenheimer, J. R., Snyder, H.:  Phys. Rev. \textbf{56}, 455 (1939)
\item[8.]Vaidya, P. C.: Nature \textbf{171}, 260 (1953)
\item[9.] Misner, C. W.: Phys. Rev. B \textbf{137}, 1350 (1965)
\item[10.]Lindquist, R. W., Schwartz, R. A., Misner, C. W.:  Phys. Rev. B \textbf{137}, 1364 (1965)
\item[11.]Herrera, L., Santos, N. O.: Phys. Rev. D \textbf{70}, 084004 (2004a)
\item[12.]Herrera, L., Di Prisco,  A., Martin, J., Ospino, J.:  Phys. Rev. D \textbf{74}, 044001 (2006)
\item[13.] Mitra, A.: Phys. Rev. D \textbf{74}, 024010 (2006c)
\item[14.] Mitra, A.: Mon. Not. Roy. Astron. Soc. Lett. \textbf{367}, L66 (2006d)
\item[15.] Tewari, B. C.: Astrophys. Space Sci. \textbf{149}, 233 (1988)
\item[16.] Tewari, B. C.: Indian J. Pure Appl. Phys. \textbf{32}, 504 (1994)
\item[17.] Tewari, B. C.:  Astrophys. Space Sci. \textbf{306}, 273 (2006)
\item[18.] Tewari, B. C.: Radiating Fluid Balls in General Relativity, VDM Verlag (2010)
\item[19.] Pant, N., Tewari, B. C.:  Astrophys. Space Sci. \textbf{331},  645 (2011)
\item[20.]Glass, E. N.: Phys. Lett. A \textbf{86}, 351 (1981)
\item[21.]Santos, N. O.:  Mon. Not. R. Astron. Soc. \textbf{216}, 403 (1985)
\item[22.] de Oliveira, A. K. G., Santos, N. O., Kolassis, C. A.: Mon. Not. R. Astron. Soc. \textbf{216}, 1001 (1985)
\item[23.] de Oliveira, A. K. G., Santos, N. O.: Astrophys. J \textbf{312},  640 (1987)
\item[24.]Bonnor, W. B., de Oliveira, A. K. G., Santos, N. O.: Phys. Rep. \textbf{181}, 269 (1989)
\item[25.]Banerjee, A., Chaterjee, S., Dadhich, N.: Mod. Phys. Lett. A \textbf{35}, 2335 (2002)
\item[26.]Govinder, K. S., Govender, M.: Gen. Relativ. Gravit. \textbf{44}, 147 (2012)
\item[27.] Maharaj, S. D., Govender, G., Govender, M.: Gen. Relativ. Gravit. \textbf{44}, 1089 (2012)
\item[28.]Herrera, L., Di Prisco, A., Martin, J., Ospino, J., Santos, N. O., Triconis, O.: Phys. Rev. D \textbf{69}, 084026 (2004b) 
\item[29.] Herrera, L., Di Prisco, A., Carot, J.: Phys. Rev. D \textbf{76}, 0440012 (2007)
\item[30.] Herrera, L., Di Prisco, A., Ospino, J., Fuenmayor, E., Triconis, O.: Phys. Rev. D \textbf{79}, 064025 (2009)
\item[31.]Naidu, N. F., Govender, M.: J. Astrophys. Astron.\textbf{ 28}, 167 (2007)
\item[32.] Sarwe, S., Tikekar, R.: Int. J. Mod. Phys. D \textbf{19}, 1889 (2010)
\item[33.]Ivanov, B. V.: Gen. Relativ. Gravit. \textbf{44}, 1835 (2012)
\item[34.]Pinheiro, G., Chan, R.: Gen. Relativ. Gravit.\textbf{ 45},  243 (2013)
\item[35.] B. C. Tewari: Astrophys. Space Sci. \textbf{3}, 1141 (2012)
\item[36.]Tewari, B. C.: Gen. Relativ. Gravit. \textbf{45}, 1547 (2013)
\item[37.]Tewari, B. C., Charan, Kali: Astrophys. Space Sci. \textbf{351}, 613 (2014)
\item[38.] Tewari, B. C., Charan, Kali: Journal of Modern Physics. \textbf{6}, 453 (2015a)
\item[39.] Herrera, L., Santos, N. O.: Physics Reports \textbf{286}, 53 (1997)
\item[40.] Herrera, L., Di Prisco, A., Hernandez-Pastora, J. L., Santos, N. O.: Physics Letters A, \textbf{237}, 113 (1998)
\item[41.]Sharma, R., Tikekar, R.: Gen. Relativ. Gravit. \textbf{44}, 2503 (2012)
\item[42.] Nguyen, P. H., Pedraza, J. F.: Phys. Rev. D \textbf{88},  064020 (2013)
\item[43.]Nguyen, P. H., Lingam, Manasvi: MNRAS, \textbf{436}(3),  2014 (2013)
\item[44.]Sharif, M., Abbas, G.: J. Phys. Soc. Jpn. \textbf {82},  034006 (2013)
\item[45.] Abbas, G.: Astrophys. Space Sci. \textbf{350}, 357 (2014)
 \item[46.]Tewari, B. C., Charan, Kali: Astrophys. Space Sci. \textbf{357}, 107 (2015b)
\item [47.] Ivanov, B.V.: Int. J. Mod. Phys. D \textbf{20}, 319 (2011)
\item [48.] Ivanov, B.V.: Astrophys. Space Sci. \textbf{361}, 18 (2016)  
\item[49.]Glass, E. N.: J. Math. Phys. \textbf{20}, 1508 (1979)
\item[50.] Cahill, M. E., McVittie, G. C.: J. Math. Phys. \textbf{11}, 1382 (1970)
\item[51.]Misner, C. W., Sharp, D. H.: Phys. Rev. B \textbf{136}, 571 (1964)
\item[52.]Israel, W., Stewart, J.:  Ann. Phys. \textbf{118},  341 (1979)
\item[53.]Maartens, R.: Class. Quantum Grav. \textbf{12}, 1455 (1995)
\item[54.]Martinez, J.: Phys. Rev. D \textbf{53},  6921 (1996)
\item[55.]Schwarzschild, M.: Structure and Evolution of Stars. Dover New York (1958)
\item[56.]Misner, C. W., Sharp, D. H.: Phys. Lett. \textbf{56}, 455(1965)

\end{enumerate}
\end{document}